\documentclass[%
 aip,jcp
 amsmath,amssymb,
preprint,%
]{revtex4-1}

\usepackage{graphicx}
\usepackage{dcolumn}
\usepackage{bm}

\usepackage[utf8]{inputenc}
\usepackage[T1]{fontenc}
\usepackage{mathptmx}

\begin{document}

\preprint{AIP/123-QED}

\title{Role of lithium intercalation in fluorine-doped tin oxide thin films: Ab-initio calculations and experiment}

\author{Israel Perez}
 \email{israel.perez@uacj.mx}
 \affiliation{National Council of Science and Technology (CONACYT)-Department of Physics and Mathematics, Institute of Engineering and Technology, Universidad Autonoma de Ciudad Juarez, Av. del Charro 450 Col. Romero Partido, C.P. 32310, Juarez, Chihuahua, Mexico }
 \affiliation{Department of Natural and Exact Sciences, CU Valles, Universidad de Guadalajara, Carr. a Guadalajara km. 45.5, C.P. 46600, Ameca, Jalisco, Mexico}

\author{Victor Sosa}%
\affiliation{Applied Physics Department, CINVESTAV Unidad M\'erida, km 6 Ant. Carretera a Progreso, A.P. 73, C.P. 97310 Merida, Yucatan, Mexico}%

\author{Fidel Gamboa}
\affiliation{Applied Physics Department, CINVESTAV Unidad M\'erida, km 6 Ant. Carretera a Progreso, A.P. 73, C.P. 97310 Merida, Yucatan, Mexico}%

\author{Jose Luis Enriquez-Carrejo}
\affiliation{Department of Physics and Mathematics, Institute of Engineering and Technology, Universidad Autonoma de Ciudad Juarez, Av. del Charro 450 Col. Romero Partido, C.P. 32310, Juarez, Chihuahua, Mexico}%
\author{Juan Carlos Mixteco Sanchez}
\affiliation{Department of Natural and Exact Sciences, CU Valles, Universidad de Guadalajara, Carr. a Guadalajara km. 45.5, C.P. 46600, Ameca, Jalisco, Mexico}

\date{\today}

\begin{abstract}
Using a combination of experimental techniques and density functional theory (DFT)  calculations, the influence of lithium insertion on the electronic and electrochemical properties of fluourine-doped SnO$_2$ (FTO) is assessed. For this purpose, we investigate the electrochromic behavior of commercial FTO electrode embedded in a solution of lithium perclorate (LiClO$_4$). The electrochromic properties are evaluated by UV-VIS spectroscopy, cyclic voltammetry, and chronoamperometry. These tests show that FTO exhibits electrochromism with a respectable coloration efficiency ($\eta=47.9$ cm$^2$/C @ 637 nm). DFT study indicates that lithium remains ionized in the lattice, raising the Fermi level about 0.7 eV deep into the conduction band. X-ray photoelectron spectroscopy (XPS) is used to study chemical bonding and oxidation states. XPS analysis of the Sn 3d core levels reveals that lithium insertion in FTO induces a shift of 350 meV in the Sn 3d states suggesting that lithium is incorporated into the SnO$_2$ lattice.
\end{abstract}

\maketitle

\section{\label{sec:level1}}

Electrochromism is the effect of changing the optical properties of a material upon the intercalation/deintercalation of ions induced by an applied electric field. Recently, electrochromic devices (ECDs) have gained considerable attention from the scientific community due to their potential applications as sunglasses, displays, touchscreens, and smart windows \cite{wu2018}. It is widely believed that the minimum number of layers to build an ECD cannot be less than five. These layers are arranged in sequence: two transparent conductive oxides (TCO) at the edges, making contact with two active electrochromic materials forming a sandwich with the ionic conductor in the middle. The first electrochromic material plays the role of a working electrode and the second one as a counter electrode. Researchers have made great efforts both to reduce cost and to increase efficiency of these devices \cite{cummins2000,lee2006,zhang2018}. Certainly, reducing the number of layers would be a great success because cost and production time would be substantially lowered.

Transition metal oxides (TMO) such as WO$_3$, CoO, NiO, Ta$_2$O$_5$, MoO$_3$, IrO, TiO$_2$, V$_2$O$_5$, MnO, are typically used in the fabrication of electrode layers \cite{monk2007}. On the other hand, thin layers of indium tin oxide (ITO) are commonly used as TCOs \cite{acuna2021}. Incidentally, thin films of ITO has been reported to exhibit electrochromic properties in the past. Earlier work of lithium insertion/extraction in ITO has shown discrepancies among several findings \cite{svensson1985,corradini1990}, most probably due to different preparation methods, although it is clear that optical properties are severely affected by ion insertion/extraction\cite{LIU20151374}. Other TCOs such as zinc oxide (ZnO) and fluorine-doped tin oxide (FTO) have also been applied in solid acid catalysts, ECD, supercapacitors, displays, and solar cells \cite{salameh2017,el2021,minami2005,yang2016}. In a recent paper Wang et al. \cite{wang2019} successfully built an ECD based on a FTO electrode which works both as an electric contact and a working electrode, reducing the number of layers to just three. The FTO electrode was embedded along with a lithium counter electrode in a cell filled with a LiPF$_6$/ethylene carbonate (EC) + dimethyl carbonate (DMC) + ethylmethyl carbonate (EMC) electrolyte. To the best of our knowledge, this is the first evidence of electrochromism in FTO. While this work shows that FTO exhibits the electrochromic effect with good stability, the investigation focuses on addressing the crystallography, morphology, and electrochromic behavior of the device; nevertheless, further investigation is required to elucidate the role of lithium intercalation on the chemical and electronic properties of FTO electrodes.

In order to fill this gap, here we use a combination of experimental techniques and theoretical calculations based on density functional theory (DFT). This methodology helps us to shed some light on the influence of Li-doping in the chemical and electronic properties of FTO thin films. To this end, we first evaluate the density of states (DOS) and band structure of FTO and Li-doped FTO (LiFTO). These calculations serve as a basis to gain some knowledge at the fundamental level. X-ray photoelectron spectroscopy (XPS) measurements of the valence band are also performed and analyzed along with DFT simulations of photoelectron spectroscopy. The electrochromic performance of a commercial FTO thin film embedded in 1 M lithium perclorate (LiClO$_4$) + propylene carbonate (PC)  + EC electrolyte is evaluated by UV-VIS spectroscopy, cyclic voltammetry, and chronoamperometry. Chemical states and chemical bonding of the samples are obtained by analyzing the Sn 3d and O 1s core levels from XPS measurements. 
\section{\label{sec:level2}Methods}

\subsection{Computational Methodology}
\begin{figure}[t!]
\begin{center}
\includegraphics[width=8.5cm]{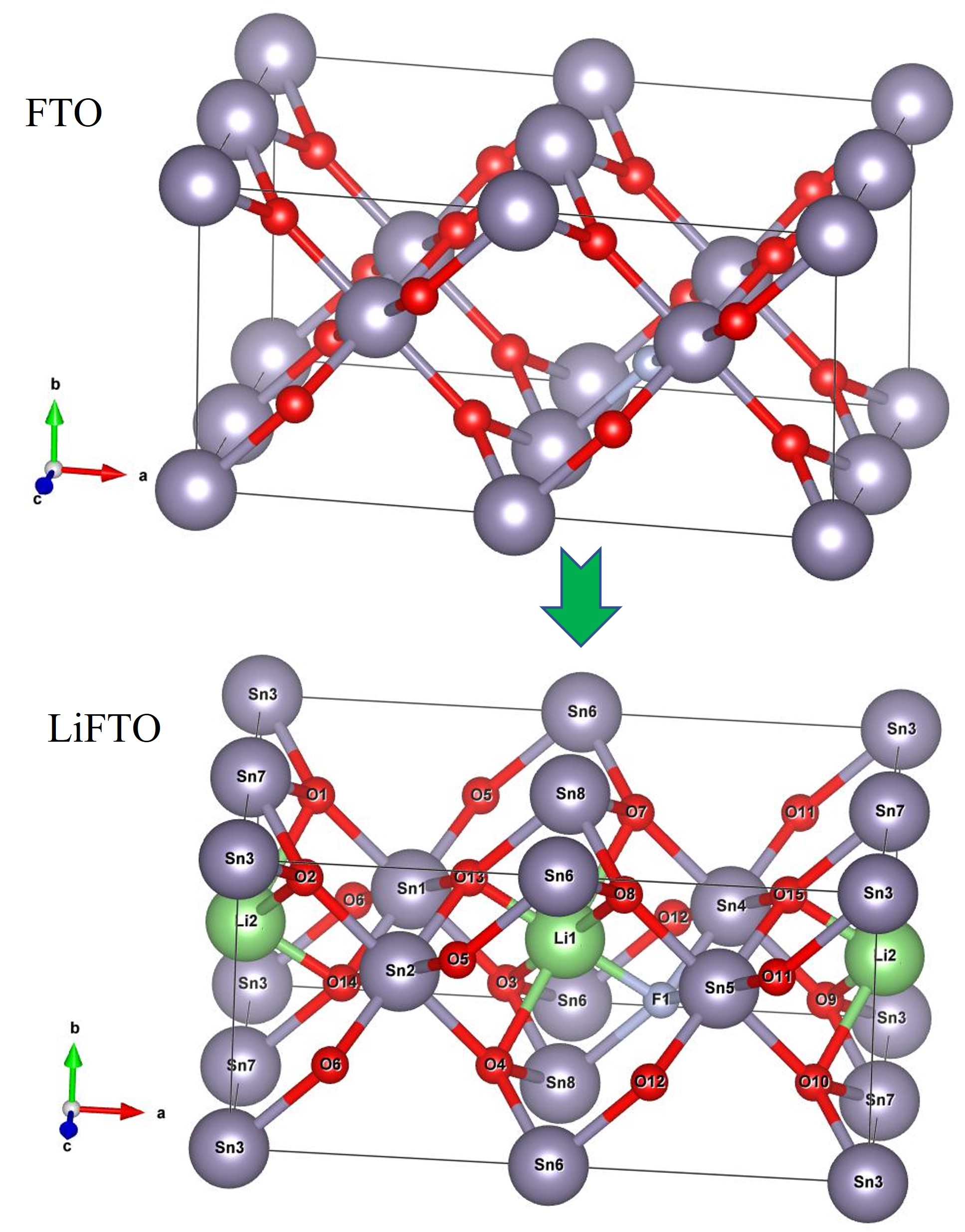}
\caption{Crystal structure of FTO and LiFTO. Atomic structure visualization provided by
the VESTA software package \cite{momma2008}}
\label{Structure}
\end{center}
\end{figure}

Density of states, band structure, and XPS valence band for FTO and LiFTO systems are computed using the full-potential linearized augmented plane wave plus localized orbitals (FP-LAPW+lo) method as implemented in \textsc{wien2k} code \cite{blaha2020}. For the exchange and correlation functional we use the Perdew-Burke-Ernzerhof (PBE) approximation \cite{perdew1996}. To model Li insertion in FTO, we start with the parent lattice of SnO$_2$ which has a rutile structure with space group P4$_2$/mnm belonging to the tetragonal system \cite{xu2009}. The lattice parameters of the optimized structure are: $a=b=4.8307$ \AA, $c=3.2421$ \AA, and $\alpha=\beta=\gamma=90^\circ$. The unit cell has two coordinate geometries with the tin atoms at the $2a$ sites (0,0,0) and (1/2, 1/2, 1/2), and four oxygen atoms at 4f sites: $(0.306a, 0.306a, 0)$, $(1/2a, 1/2a, 1/2c)$+$(-0.306a, 0.306a, 0)$. To form FTO a 2$\times$1$\times$2 supercell is generated and one O atom is replaced by a F atom, leading to the final configuration Sn$_8$O$_{15}$F (see Fig. \ref{Structure}). 

Intercalation of alkali metals in transition metal oxides may induce phase transitions as a function of impurity content. This can be avoided by keeping doping levels low \cite{perez2021}. The Li doping level is determined by several factors but the most important is certainly the number of sites available. Although the study of the electronic properties caused by substitutional doping is certainly of great interest, here we restrain ourselves to deal with interstitial doping only and leave the other case for future work. The allocation of dopants in a lattice must take into consideration the size of the interstices and the volume of the doping atoms, thus avoiding overlapping among atomic spheres. Interstitial Li positions were chosen by visual inspection of the FTO structure keeping in mind that Li atoms should respect lattice symmetry. Accordingly, two possible positions were singled out, namely: (1/2$a$,1/2$a$,1/2$c$) and (0,1/2$a$,1/2$c$). In the present study we only deal with the former (see Fig. \ref{Structure}) where only one lithium atom is added in the 2$\times$1$\times$2 supercell (labelled Li1 in the Fig.); this represents 4.2 at.\% of Li doping.

The k-space integrations are performed with a $4\times8\times6$ Monkhorst-Pack $k$-mesh based on the tetrahedron method. The cut-off energy between core and valence states is set to $-7$ Ry. The radii of the muffin tin (R$_{MT}$) spheres for the atoms (given in atomic units) are chosen so that the neighboring spheres are nearly touching. The selection of these radii allows us to avoid not only any charge leakage but also any sphere overlapping. The radii are $R_{\textrm{Sn}}=2.10$ a.u., $R_{\textrm{O}}=1.72$ a.u., $R_{\textrm{F}}=1.70$ a.u., and $R_{\textrm{Li}}=1.50$ a.u.

Structures are relaxed by first carrying out a lattice optimization and then relaxing the atomic forces which in turn define the internal atomic coordinates. This procedure is repeated until convergence is reached. Atomic forces are relaxed up to 2 mRy/Bohr Hellman-Feynman forces. 

For the expansion of the basis set, we set the product of the smallest of the atomic sphere radii $R_{MT}$ and the plane wave cutoff parameter $K_{max}$ to $RK_{max}$= 7. During the self-consistent field (SCF) cycle, the energy convergence is less than 100 $\mu$Ry/unit cell, charge convergence is less than 1 m$e$ (with $e$ the electron charge) and a force convergence of less than 1 mRy/Bohr. For higher resolution in the computation of the DOS and XPS valence band spectra, the size of the k-point grid is increased to $5\times10\times8$.

 The XPS spectra of the valence band are simulated and compared with experimental measurements. The program \textsc{PES}, also implemented in the \textsc{WIEN2k} package, is used to compute the valence band from the partial DOS of the 4d, 5p and 5s states of Sn; the 2s and 2p states of O and of F; and the 2s states of Li. To account for life-time broadening, the calculated DOS has been broadened with a Lorentzian distribution with a full width at half maximum (FWHM) of 0.01 eV and a Gaussian distribution with FWHM equal to 0.07 eV. The program multiplies the DOS by the total cross section and the inverse of the fraction of the charge located inside the atomic spheres as explained elsewhere\cite{bagheri2019}.

\subsection{Experimental}
Two commercial FTO/glass thin films with an area of 2.5 cm$\times$2.5 cm are used for the experimental part; they were acquired from MTI corporation Richmond Ca. USA . The thickness of the glass is 2.2 mm and that one of FTO film is about 200 nm. The sheet resistance of the films is $6-9\; \Omega/\square$. One sample is exposed to electrochemical tests, as described in the next subsection, while the other is left as a reference. Henceforth we shall call the former the lithiated sample (LiFTO) and the latter the as-received (AR) sample.

\subsection{Characterization}
A Thermo Scientific K-Alpha XPS spectrometer with an Al K$_{\alpha}$ X-ray source set to 12 kV and 40 W is used to study the chemical properties of the films with and without lithium. For the XPS scans, we use steps of 0.1 eV; the beam spot has a diameter of 400 $\mu$m and makes an angle relative to the sample surface of 30$^\circ$. Chemical properties are assessed measuring the Sn 3d, O 1s and C 1s core levels. The chemical states and atomic bonding are determined by deconvolution of the C 1s, Sn 3d, and O 1s spectra using Tougaard background and Voigt functions as implemented in the CasaXPS software (version 2.3.19PR1.0). Sn spectra are deconvoluted using doublets with a spin-orbit splitting of 8.42 eV. XPS valence band spectra for both samples are also recorded and compared with DFT simulations. Accordingly, a linear background is used and subtracted from the raw data.

Transmittance of both samples is measured with an UV-VIS system VWR model UV-1600 in a wavelength interval from 300 nm to 1000 nm. To determine the electrochromic characteristics, cyclic voltammetry (CV) and chronoamperometry (CA) are carried out using a CorrTest CS350 electrochemical workstation in a three electrode set up. A platinum electrode is used as a counter electrode and a Ag/AgCl$_{\textrm{KOH}}$ electrode as a reference electrode. One of the samples is used as a working electrode and the three electrodes are embedded in a cubic optoeletrochemical cell (125 cm$^3$) filled with the electrolyte. The electrolyte is a solution of 1 M of LiClO$_4$ dissolved in PC and EC. To eliminate moisture, the lithium salts were previously heated up to 140 $^\circ$C for 1 h.

Transmittance in CA measurements is obtained at a fixed wavelength of 637 nm, using a high sensitivity sensor Pasco (model CI-6604) connected to a Pasco interface UI-500. CA data are obtained in the range from $-3$ V to 3 V with steps of 20 s each. CV measurements are realized for the same voltage interval with a scan rate of 50 mV/s. 

\section{Results and Discussion}
\label{sec3}
\subsection{Electronic Structure}
We first discuss the theoretical outcome. The optimized structure of FTO yields lattice parameters $a=b=5.1015\,$\AA, $c=3.0815\,$\AA; corresponding to a cell volume of $80.1969\,$ \AA$^3$. Compared to the cell volume of SnO$_2$ there is an increase of 6\%. For LiFTO the parameters are: $a=b=5.1666\,$\AA, $c=6.2416\,$\AA, and a cell volume of $83.3073\,$\AA$^3$; representing, with respect to FTO, a 3.8\% cell expansion as a consequence of lithium insertion. Cell expansion is expected in fluorine-doped SnO$_2$ \cite{salameh2017} and in lithium-doped metal oxides \cite{wang2019,perez2021}. In FTO Li insertion exerts chemical pressure on the initial atomic positions deforming the lattice; such structural variations in turn affect the electronic structure. 

\begin{figure}[t!]
\begin{center}
\includegraphics[width=8.5cm]{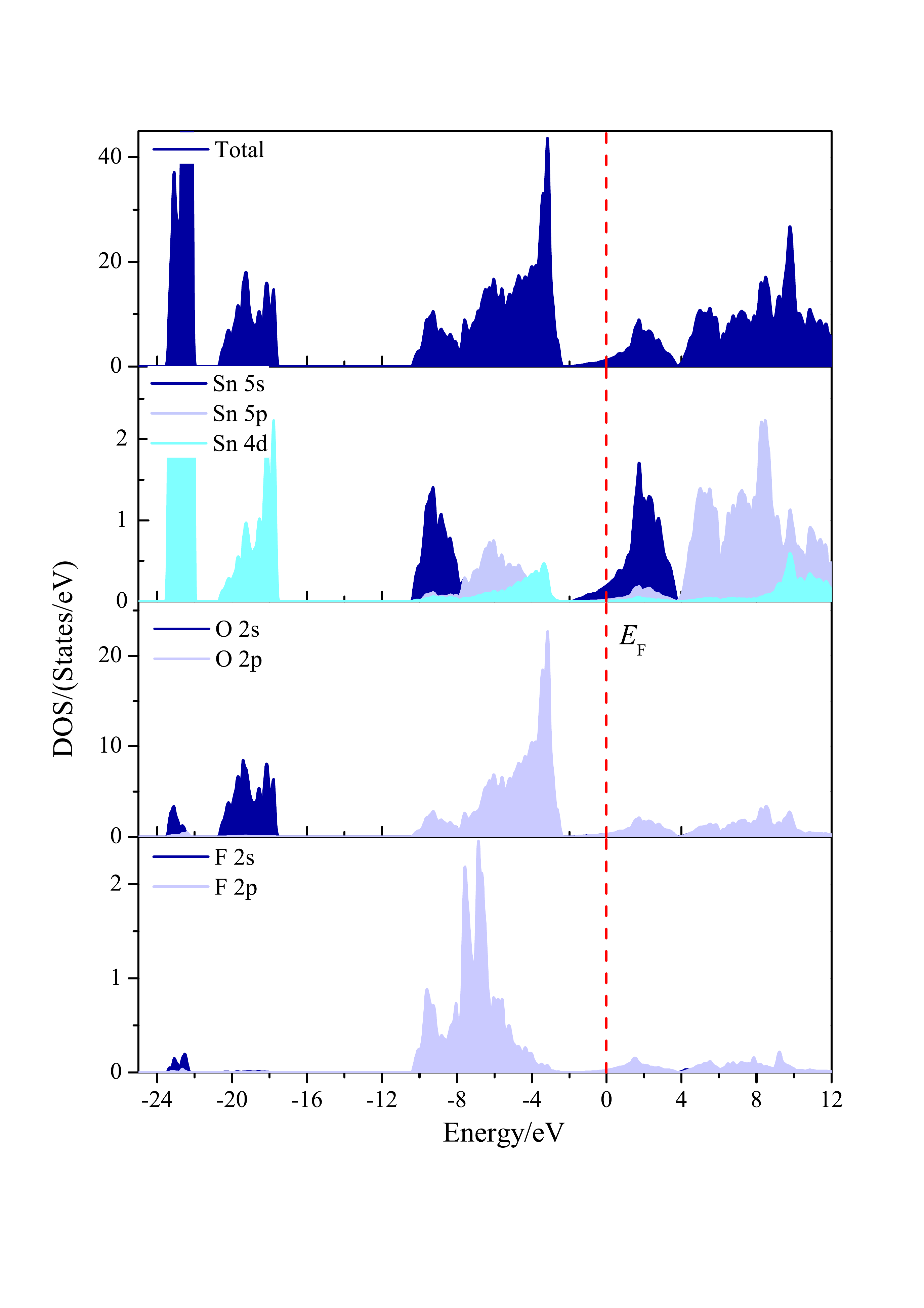}  
\caption{Total and projected density of states of FTO}
\label{DOSFTO}
\end{center}
\end{figure}

\begin{figure}[t!]
\begin{center}
\includegraphics[width=8.5cm]{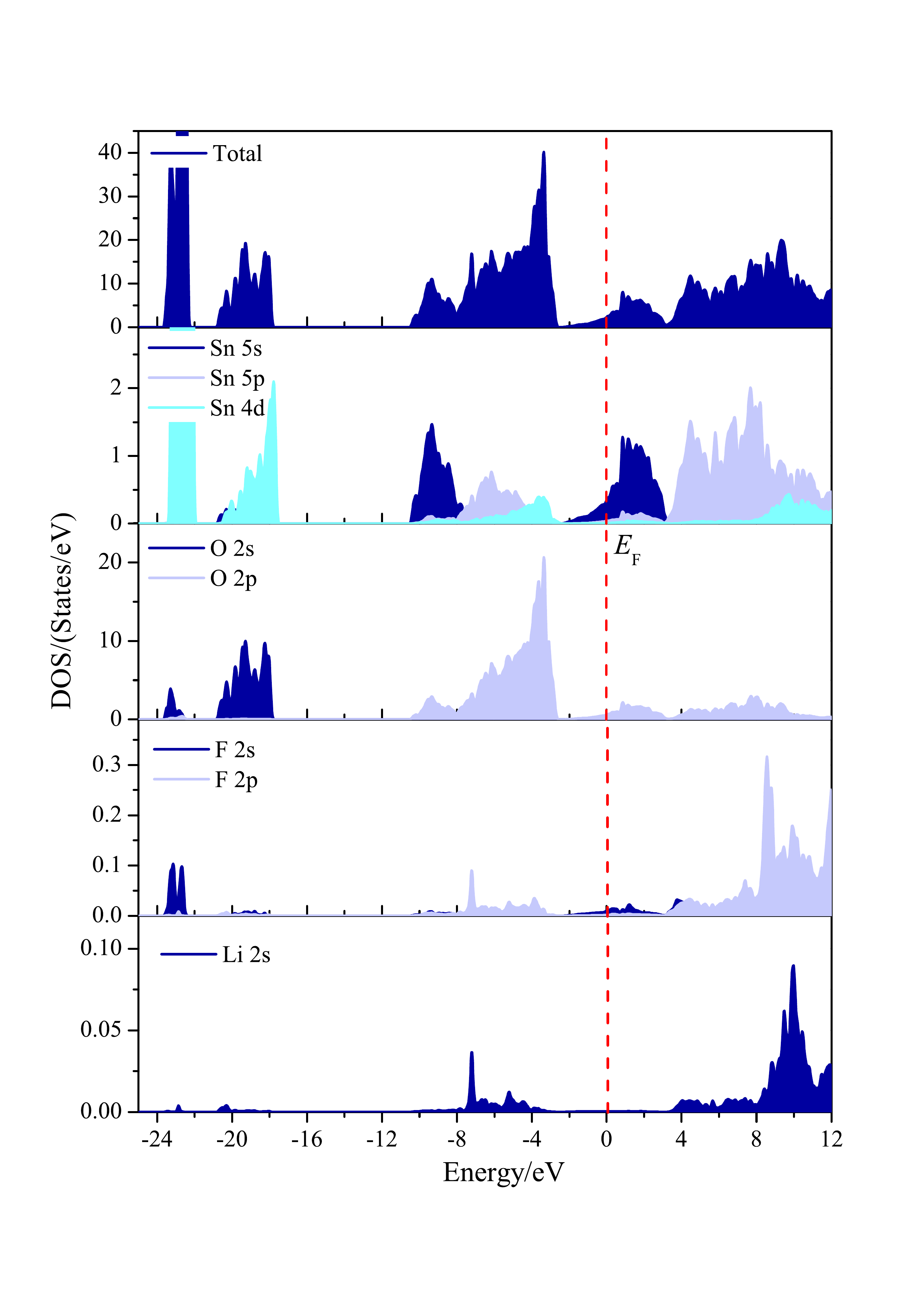}  
\caption{Total and projected density of states of LiFTO}
\label{DOSLiFTO}
\end{center}
\end{figure}

In order to study the influence of lithium intercalation in the electronic properties of FTO, the DOS and the band structure are obtained from DFT quasiparticle eigenenergies. Figures \ref{DOSFTO} and \ref{DOSLiFTO} show the total and projected DOS for the two systems. The results for FTO are in good agreement with a previous report \cite{xu2009}. The vertical line is the Fermi level ($E_{\textrm{F}}$) that divides the occupied states from the unoccupied states, showing that both systems behave as conductors. In both cases, most of the valence band is dominated by O 2p states; except for a high contribution of Sn 4d states in the very low energy region. Both systems display a prominent peak between $-2.5$ eV and $-4$ eV which has O 2p bonding or nonbonding character. The region between $-4$ eV and $-8$ eV of FTO is caused by a $pp\sigma$ bonding orbital involving Sn 5p, O 2p, and F 2p states. For LiFTO this region is mainly due to hybridization of Sn 5p and O 2p orbitals since F 2p states are highly suppressed. The region from $-8$ eV to $-12$ eV of FTO is a bonding state produced by the strong hybridization of Sn 5s, O 2p, and F 2p orbitals which strongly suggests ionic bonding between Sn and F atoms. On the other hand, the presence of Li in FTO distorts the lattice and induces an additional effect on the F 2p states whose DOS is highly suppressed in this region, reducing the Sn $5$s$-$F 2p interactions. The reason is that Li cations exert chemical pressure and displace the fluorine atoms from their initial equilibrium positions.
Moreover, it is evident that Li 2s states have a negligible contribution, nonetheless, these states strongly hybridize with p states and Sn 4d states, donating their electrons to the conduction band and showing that the Li atom remains ionized in the lattice. Therefore, Li in LiFTO plays the role of an electron donor.

\begin{figure}[t!]
\begin{center}
 \includegraphics[width=8.5cm]{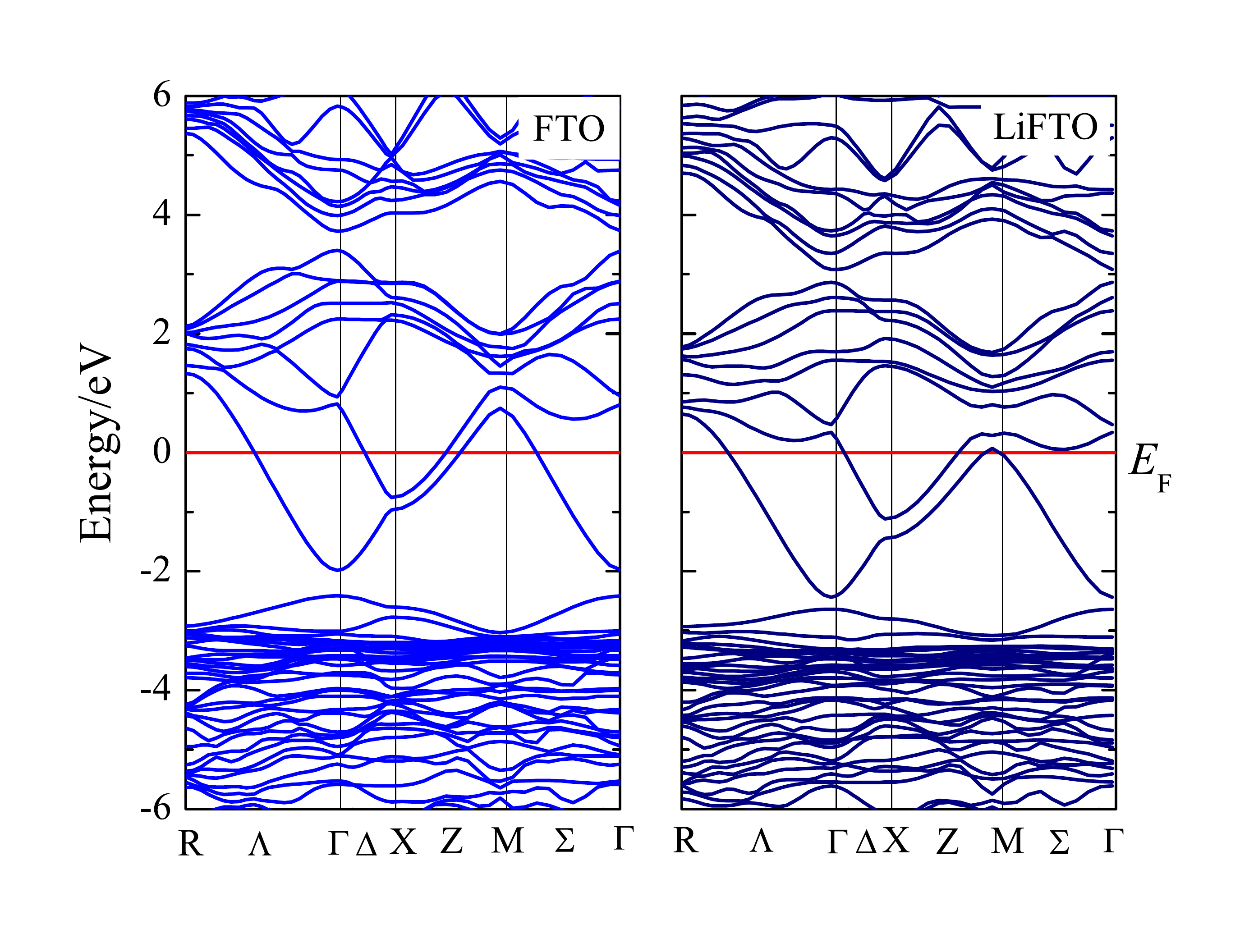} 
\caption{Band structure of FTO and LiFTO. Red line denotes the Fermi level. Vertical lines project high symmetry points in the first Brillouin zone.}
\label{OptimiFTOLiFTOBS}
\end{center}
\end{figure}

The band structure of FTO and LiFTO are depicted in Fig. \ref{OptimiFTOLiFTOBS} along with high symmetry lines in the first Brillouin zone. By comparing the band structures of both systems, it is seen that, overall, the band distribution remains the same. We note, however, that lithium intercalation has the effect of rising the Fermi level about 0.7 eV deep into the conduction band. It is well known that SnO$_2$ is an n-type  semiconductor with a wide direct band gap of $(3.6-4.0)$ eV \cite{reimann1998}. After doping with fluorine, the material turns into a semimetal with sheet resistance below 20 $\Omega/\square$ \cite{salameh2017}. Fluorine acts in the SnO$_2$ as a donor impurity and thus causes degeneracy. The semimetallic character prevails after lithium insertion. Electrons from Li 2s states are incorporated into the system and more levels are required to allocate the extra electrons, therefore rising the Fermi level deep into the conduction band. This also has the effect of narrowing the band gap of FTO, indicating that there is a small energy shift of the optical absorption edge of LiFTO compare to that of FTO.

\begin{figure}[t!]
\begin{center}
 \includegraphics[width=8.5cm]{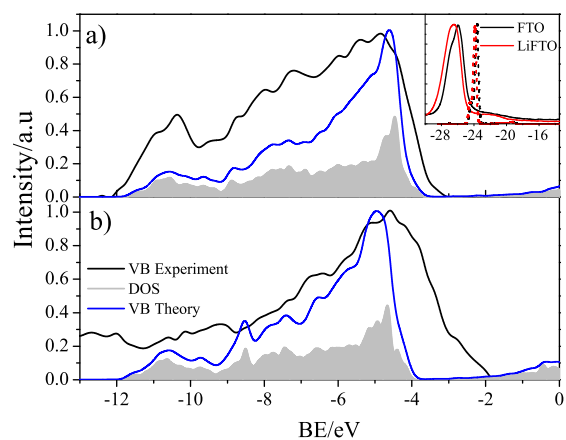} 
\caption{XPS valence band spectra for the AR sample (a) and LiFTO sample (b). Experiment in black, simulation in blue, and DOS in grey. Inset shows the low energy region for the Sn 4d states. Solid lines correspond to measurements and dashed lines to theoretical predictions.}
\label{VBFTOLiTO}
\end{center}
\end{figure}

To further elucidate the effect of lithium on the electronic properties, the XPS valence band spectra are analyzed (see Fig. \ref{VBFTOLiTO}) for both samples (black spectra) and compared with DFT calculations of photoelectron spectra (blue spectra). As a reference we also add the total DOS (gray spectra). Valence band spectra were normalized for comparison. To match the experimental spectra, calculated spectra of both systems were manually shifted 1.3 eV towards lower energies. This shift is justified by the fact that eigenvalues from DFT do not represent excited states. The measurements of the AR sample compare well with an earlier investigation by Mart\'inez et al. \cite{martinez2006}. They showed that the feature around $-10.6$ eV is due to fluorine content. Our simulation predicts it well as a result of hybridization of Sn 4d, O 2p, and F 2p states. By contrast, it is slightly suppressed in the LiFTO sample although it is still predicted by the theory. It is worth mentioning that this is the first report to have measured the valence band spectra of Li-doped FTO by XPS; so DFT calculations are useful for the interpretation of this information. The theoretical results for LiFTO reveal a prominent peak at $-8.5$ eV which is caused by hybridization of Sn, F, O p states and Li s states, however, in the experiment this feature is barely seen. We assume that this may be related to lithium content, since lithium displaces fluorine atoms from their equilibrium positions changing the local charge density. 

As an additional feature, the theory indicates a binding energy shift of $\sim0.3$ eV for the valence band of LiFTO relative to that one of FTO. Compared with the measurements, this shift is not noticeable in the high energy region (due to peak broadening) although it is evident at low energies, particularly for the Sn 4d peak around $-26$ eV (see inset in the same figure). The observed shift amounts to 0.5 eV (solid lines) although the predicted peaks (dashed lines) appear about 2 eV higher in energy. The discrepancy in the peak positions can be traced back to the failure of DFT in computing the correct band structure of many semiconductors and strongly electron correlated materials. It is well known that the local density approximation as well as the generalized gradient approximations, such as the PBE, underestimate the magnitude of the band gap due to a discontinuity occurring in the functional derivative of the exchange correlation energy functional \cite{yakovkin2007}. Better results might be obtained using a higher level of theory such as the GW or the Bethe-Salpeter approximation. Unfortunately, these resources are beyond our computation capabilities. Despite this, we think that our DFT simulations give a good qualitative and semiquantitative description of experimental data.

\subsection{Optical and electrochromic properties}
\begin{figure}[b!]
\begin{center}
 \includegraphics[width=8.5cm]{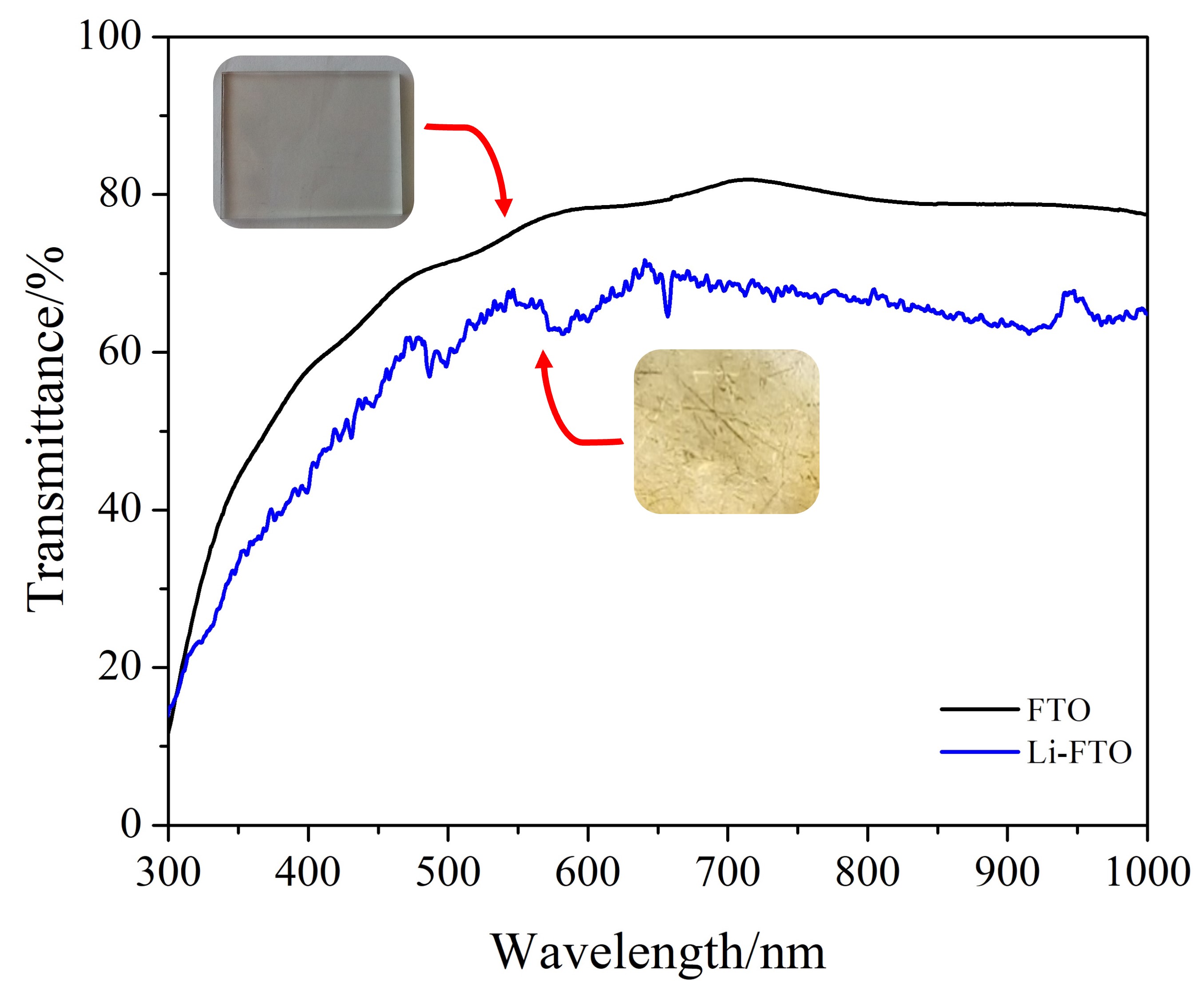} 
\caption{Transmittance of FTO with and without lithium insertion. For the latter a bias of $-2$ V was used.}
\label{TransFTOLiTO2}
\end{center}
\end{figure}
\begin{figure*}[t!]
\begin{center}
\includegraphics[trim=0cm 3cm 1cm 0cm, clip=true, width=17cm]{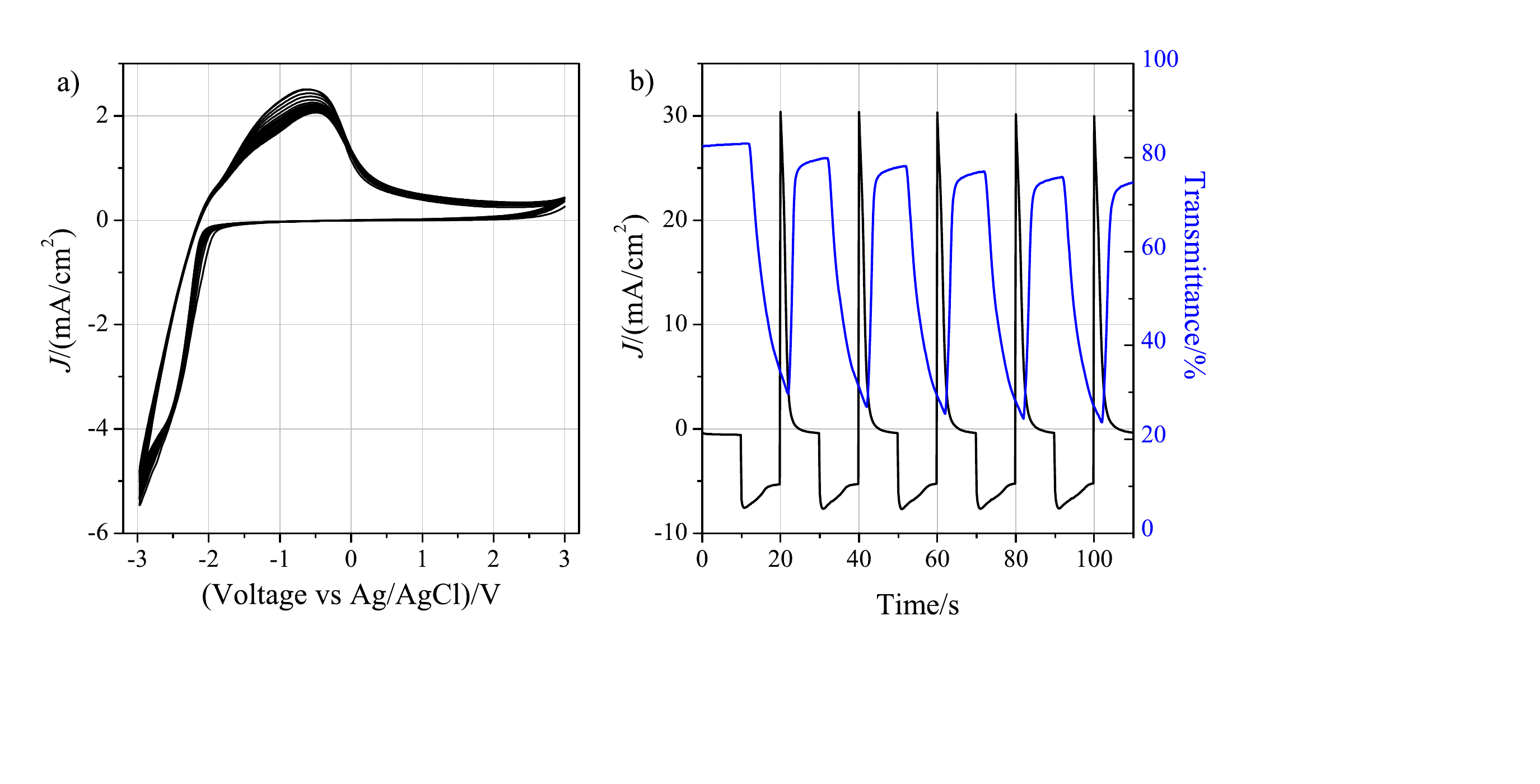} 
\caption{Cyclic voltammetry (a) and chronoamperometry (b) measurements. Curves of transmittance vs time are also included and contrasted to CA measurements.}
\label{CACVFTO}
\end{center}
\end{figure*}
In Fig. \ref{TransFTOLiTO2} the transmissivity of the AR and LiFTO samples is displayed. The lithiated data was taken at bias of $-2$ V for 2 min. The AR sample looks transparent to the naked eye  while the lithiated sample turned into a yellowish color in the zone immersed in the electrolyte (see insets in the same figure). It is seen that the transmissivity for the AR sample (black curve) reaches a maximum of about 82\% at 700 nm and remain almost constant for higher wavelengths, in agreement with previous work \cite{wang2019,ren2018}. On the other hand, the LiFTO sample presents a maximum of 71\% at 630 nm. The fluctuations along the curve can be attributed to lithiation effects since, as can be seen in the image of the sample, the darkening is not homogeneous and the same zone may respond differently to wavelength variations. 

CV and CA measurements are conducted to assess the electrochromic properties; data is displayed in Fig. \ref{CACVFTO}. During the CV measurements, the film changed from transparent to dark yellow in the cathodic potential (from positive to negative potentials) by intercalation of Li$^+$ ions from the electrolyte solution into the FTO film. Bleaching occurs in the opposite direction, extracting Li$^+$ ions from the film and thus presenting cathodic electrochromism. Reduction peaks are absent in cathodic potential, while prominent oxidation peaks occur in anodic potentials (from negative to positive potentials) around $-0.6$ V. As the number of cycles increases the intensity of the oxidation peak slightly decreases. We think that the oxidation peak might be related to the oxidation of metallic Sn.

The coloration efficiency for a given wavelength ($\lambda_o$) is defined as 
\begin{equation}
\eta=\frac{\Delta OD (\lambda_o)}{Q_x},
\label{col}
\end{equation}
where $Q_x$ is the inserted/extracted charge, 
\begin{equation}
\Delta OD= \ln \frac{T_b(\lambda_o)}{T_c(\lambda_o)},
\end{equation} 
is the optical density, $T_b$ and $T_c$ are the transmissivities in the bleached and colored states, respectively. We note that for the computation of the coloration efficiency we use the inserted charge ($Q_{in}$), although the extracted charge ($Q_{out}$) can also be used. These are determined from the CA measurements by integrating the area under the curve as an average over the cycles shown in the figure. $Q_{in}$ is associated to the colored state while $Q_{out}$ is the charge related to the bleached one. The area of the film immersed in the electrolyte was about 4.2 cm$^2$. 

By comparing the transmittance curves with the CA measurements, one realizes that the colored(bleached) states take place for negative(positive) currents (or equivalently negative voltages). In the bleached state, the average transmissivities are $T_b=78.1$\% and $T_c=26.3$\%. Accordingly, it is found that $Q_{in}=62.7$ mC/cm$^2$ and $Q_{out}=42.02$ mC/cm$^2$. This gives a reversibility of $K=\frac{Q_{out}}{Q_{in}}=67$\% and $\eta=47.9$ cm$^2$/C which compares well with the values reported for NiO (50 cm$^2$/C @ 550 nm) and WO$_3$ (54.8 cm$^2$/C @ 633 nm)\cite{mahmoudi2016,acuna2021}. If the Li intercalation/deintercalation process is truly reversible, $K$ must be 1, however it is clear that more than 30\% of the charge gets trapped in the FTO lattice. So, in order to have high durabilities, high reversibilities are desired.

\subsection{Chemical properties}
\begin{figure*}[t!]
\begin{center}
 \includegraphics[width=17cm]{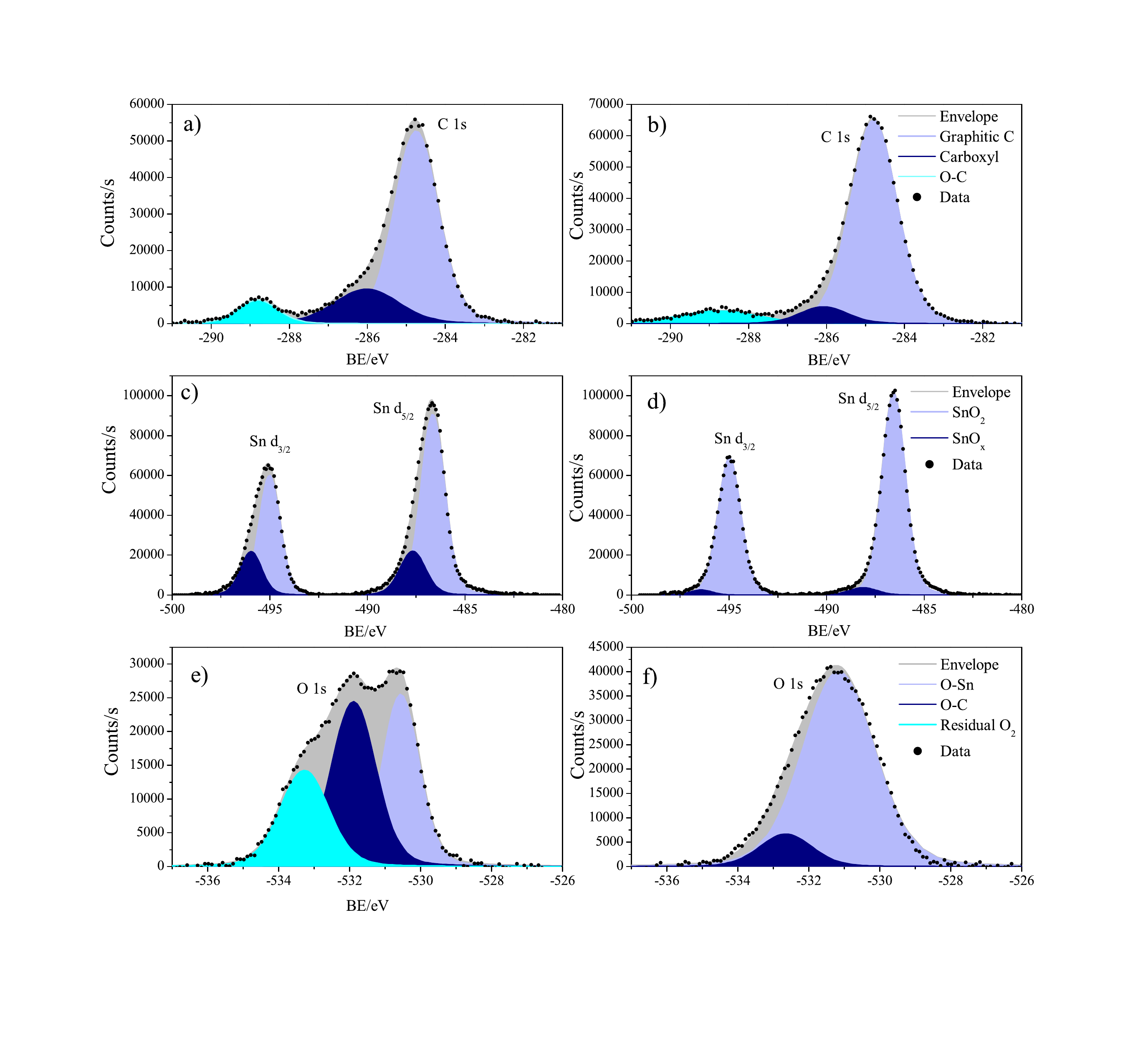} 
\caption{XPS spectra for C 1s, Sn 3d, and O 1s core levels. Left column for as-received FTO and right one for LiFTO.}
\label{XPSFTOComp2}
\end{center}
\end{figure*}
The chemical properties of both the as-received FTO and LiFTO samples are evaluated by XPS analysis. The effect of lithium insertion in the chemical properties of FTO are shown in Fig. \ref{XPSFTOComp2}. Li intercalation in FTO was induced by dipping the sample in the electrolyte solution and applying $-2$ V bias for 2 min. Samples were then placed in the XPS load-lock chamber at a base pressure of 10$^{-7}$ mbar. The C 1s core level was taken to assess the level of contamination in the samples [see Fig. \ref{XPSFTOComp2}(a) and (b)]. The signal was deconvoluted into three Gaussian peaks; for both samples they are located at 284.7 eV, 286.0 eV, and 288.8 eV. These peaks are typical of an air exposed surface. The low binding energy peak is attributed to graphitic carbon, the medium energy one is related to carbon oxide (most likely CO$_2$) or hydroxide, and the high energy one can be assigned to either carboxyl groups or to a C-F group \cite{moulder1992,Shiratsuchi_1992,kuo1992,pollard1999}. Nevertheless, the peak of fluorine F$^-$ (not shown) shows up at 685.0 eV, if we were to attribute the high binding energy peak to a C-F group, the fluorine peak would have to appear shifted to higher energies (about 2 eV) but this is not the case. We therefore rule out this possibility.

To resolve the effect of Li$^+$ cations introduced into the FTO structure, the XPS spectra of both samples are presented in Fig. \ref{XPSFTOComp2}(c) and (d), respectively. Both samples exhibit the spin-orbit doublet corresponding to the 3d$_{5/2}$ and 3d$_{3/2}$ suborbitals. For the AR sample the mean peaks are located at 486.91 eV and 495.33 eV. The peak splitting is 8.42 eV which is in good agreement with the literature for SnO$_2$ and FTO \cite{moulder1992,martinez2006}. Their peak fitting was achieved with two doublets (light blue and dark blue in the figure). The first one (light blue) is attributed to Sn-O bonding in the Sn$^{4+}$ state \cite{moulder1992}. The second one (dark blue) can be associated to Sn suboxides (SnO$_x$) or to energy losses \cite{LIU20151374,swallow2018}. Earlier investigations\cite{Shiratsuchi_1992,LIU20151374} on induced reduction of ITO and FTO reported an additional peak at lower binding energies (around 484.5 eV) which was assigned to metallic Sn. In our case, we do not observe this additional doublet, indicating that there are no signs of FTO reduction; in agreement with CV measurements. When Li is introduced, the peaks shift 0.35 eV to higher energies, this effect can be due to the difference in electronegativies between Li and Sn atoms (0.98 and 1.96, respectively) \cite{qiang2020}. 

The O 1s core levels for FTO [Fig. \ref{XPSFTOComp2}(e)] can be deconvoluted into two main Gaussian peaks with center lines located at 530.56 eV (light blue peak) and 531.89 eV (dark blue peak), where the lower binding energy peak is related to O-Sn bond in O$^{2-}$ oxidation state and the higher one emerging from chemisorbed oxygen as residual oxygen (O$_2$) or OH species as surface contamination \cite{ke2015,perez2018}. These two peaks also show up for the sample with lithium [Fig. \ref{XPSFTOComp2}(f)], however due to lithium incorporation into the lattice, they are shifted to higher energies: 531.11 eV and 532.57 eV, respectively. The additional shoulder (cyan peak) in FTO at 533.3 eV has been observed in Ta$_2$O$_5$ and SnO$_2$ compounds and has been identified as C-O binding due to carbon oxide contamination as we discussed above \cite{pollard1999,perez2019,ren2018}. 

\section{Summary and conclusions}
\label{sec4}
We have combined first principles calculations with experimental techniques (UV-VIS, CA, CV, and XPS) to study the electronic structure, electrochromic behavior and chemical bonding of commercial Li-doped FTO thin films. The present investigation sheds light on the basic aspects of Li intercalation in FTO. DFT calculations demonstrate that Li remains ionized in the FTO lattice and that Li atoms provide electrons to the system. Relative to FTO, electron doping raises the Fermi level higher into the conduction band. The valence band spectrum was analyzed by XPS and DFT simulations of photoelectron spectroscopy. Save for a binding energy offset of 1.3 eV, theoretical predictions are in good agreement with observations for FTO. Simulation shows that its valence band spectrum is composed of Sn 4d, 5s, 5p states, O and F 2p states, and O and F 2s states. An additional feature is incorporated into the spectrum by Li 2s states that is not observed in the measurements for LiFTO. However, the theory predicts a small energy shift of 0.3 eV of the valence band of LiFTO relative to that one of FTO; we believe that the observed shift of 0.5 eV can be due to Li incorporation. The discrepancy may be due to the fact that energy eigenvalues from DFT are not excited states and higher levels of theory are needed to refine the calculations. 

Electrochemical techniques were used to characterize the electrochromic behavior of commercial films of FTO. Quantitative analysis of electrochemical tests demonstrates that FTO possesses a respectable coloration efficiency ($\eta=47.9$ cm$^2$/C) which matches those from most popular electrochromic materials (NiO and WO$_3$). It is thus concluded that the electrochemical properties of FTO can be exploited to reduce the number of layers in ECDs, reducing the processing time and cost of these devices. Finally, XPS analysis of the Sn 3d core levels for FTO and LiFTO reveals a 350 meV-shift attributed to Li intercalation in FTO, supporting the view that Li is fully incorporated into the SnO$_2$ lattice as predicted by DFT simulations.  

\section*{Author's contribution}
Israel Perez conceived the project, obtained the funding, carried out the DFT simulations and electrochemical tests, analyzed the data, and wrote the manuscript. Victor Sosa and Fidel Gamboa performed the XPS and optical measurements and analyzed the data. Jose Enriquez obtained the funding and analyzed the data. Juan Mixteco analyzed data. All the authors reviewed, discussed and commented on the manuscript.

\begin{acknowledgments}
The authors gratefully acknowledge the support from the National Council of Science and Technology (CONACYT) Mexico under project 3035 and partial support from the Universidad Autonoma de Ciudad Juarez, Mexico through project PIVA 334-18-12. We thankfully acknowledge the computer resources, technical advise, and support provided by Laboratorio Nacional de Inform\'atica (LANTI-UACJ), special thanks to Oscar Ruiz and William Cauich for technical support during the computational and XPS sessions, respectively. 
\end{acknowledgments}

\section*{Author Declaration}
The authors have no conflicts to disclose.

\section*{DATA AVAILABILITY Statement}
The data that support the findings of this study are available from the corresponding author upon reasonable request.

\bibliography{FTO}

\end{document}